\documentclass{article}
\begin{document}

\begin{center}
{\large\bf
PHENOMENOLOGICAL RELATIONS FOR QUARK AND NEUTRINO MIXING
ANGLES
}

\medskip
\bigskip

{\bf Yu.V. Gaponov}\footnote{e-mail: gaponov2@imp.kiae.ru}, 
{\bf V.V. Khruschov}\footnote{e-mail: khru@imp.kiae.ru}, 
{\bf S.V. Semenov}\footnote{e-mail: semenov@imp.kiae.ru}

\smallskip

RRC "Kurchatov Institute", Kurchatov Sq.1,
123182 Moscow

\end{center}

\begin{abstract}
   The  recent experimental data on mixing angles for  quarks
and  neutrino have been discussed. Quarks mixing  angles  are
calculated  in  the  phenomenological approach  of  Fritzsch-
Scadron-Delbourgo-Rupp (FSDR approach) using mass  values  of
light  and  heavy  constituent quarks.  The  neutrino  mixing
angles have been calculated with the high degree of precision
with the help of the hypothesis of quarks and neutrino mixing
angles   complementarity,  the  results   obtained   do   not
contradict with the present experimental data.
\end{abstract}

\bigskip
\smallskip

{\bf    1. Introduction}

     An  elucidation of a character of mixing of  quarks  and
neutrino  states  is an unsettled problem of  Standard  Model
(SM).  Of fundamental importance is the calculation of mixing
angles  in agreement with experimental data. There  are  many
works  devoted to this problem, in which different approaches
have  been  proposed (e.g., see reviews [1-4]),  however  the
answer  not  conclusively  obtained.  It  is  possible   that
resolution of this issue will appear after generalization  of
SM,  when  some  new  components will be included  in  for  a
description   of  phenomena  beyond  SM  such   as   neutrino
oscillations.  An  exhaustive resolution of  the  problem  of
mixing  of  quarks  and neutrino should  appear  in  a  Great
Unification Theory (GUT), but the eventual variant of GUT  is
not  available  now.  A  certain intension  about  a  further
development  of  SM  can be obtained in  studies  of  various
mutual   relations  between  mixing  angles  of  quarks   and
neutrino.  Noted phenomenological regularities in  quark  and
lepton  sectors of SM, particularly the hypothesis of  quarks
and  neutrino mixing angles complementarity, tell us about  a
possible  existence of a common physical cause of mixing  for
these particles. The facts of such kind deduced from data may
be  useful in elucidation of a way of a SM generalization and
a  subsequent GUT development. From this point of view it  is
interesting  on  the  one  hand to  perform  an  experimental
verification of phenomenological relations for mixing  angles
with  the  best  accuracy, and on  the  other  to  pursue  an
investigation  of a possible interpretation of the  relations
in  the framework of existing models, along with a dependence
of  mixing angles from other parameters, for instance, masses
of fundamental particles.

     Inasmuch as the SM is a quantum gauge chiral theory,  so
in its initial lagrangian mass terms are absent. For example,
fermionic  masses arise from Yukawa interactions  with  Higgs
fields  after  a spontaneous breaking of gauge  symmetry.  In
doing  so fermionic states of the initial lagrangian  
$f_{\alpha}'$  and fermionic  
states with definite masses $f_i$ can  be  correlated
each other with a nontrivial mixing matrix, which assumed  to
be  an  unitary one, as a rule. Under condition  that  $u-$type
quarks  and charged leptons are states with definite  masses,
the  states  of $d-$type quarks and neutrino entered  into  the
interaction  lagrangian are connected with their mass  states
by $V_{CKM}$  and $U_{PMNS}$ matrices, correspondingly:
\begin{equation}
                     d'_{\alpha L} = V_{CKM,\alpha i}d_{iL},
\end{equation}                             
\begin{equation}
                     \nu'_{\alpha L} = U^{*}_{PMNS,\alpha i}\nu_{iL},
\end{equation}                             
where $\alpha,  i = 1,..., n$. The mixing matrix in the quark  sector
$V_{CKM}$  is  named as Cabibbo-Kabayashi-Maskawa matrix  [5,  6],
whereas  the mixing matrix in the neutrino sector  $U_{PMNS}$   is
Pontecorvo-Maki-Nakagawa-Sakata matrix [7, 8].

     It  is  common knowledge that the unitary $n\times n$  matrix  is
determined  with  $n^2$ real parameters, for instance,  $n(n-1)/2$
angles  and  $n(n+1)/2$  phases.  Take  into  account  the   SM
electroweak  lagrangian structure with  currents  constructed
from  quarks,  charged leptons and neutrino,  when  fermionic
fields  are  Dirac  type  it is possible  to  eliminate  $2n-1$
phases.  For  Majorana  type neutrino fields  only  $n$  phases
connected  with Dirac type charged leptons can be  eliminated
[9].  On  this  basis $V_{CKM}$  $n\times n-$ matrix which specify  quarks
mixing in general case is determined with $n(n-1)/2$ angles and
$(n-1)(n-2)/2$  phases. However $U_{PMNS}$ $n\times n-$ matrix is determined
with the same number of parameters if neutrino are Dirac type
or  with $n(n-1)/2$ angles and $n(n-1)/2$ phases if neutrino  are
Majorana type.

     Conceptually  a  manner  of its  own  can  be  used  for
parameterization of  $V_{CKM}$ or $U_{PMNS}$ matrix. But  in  order  to
reveal  an  association between mixing mechanisms for  quarks
and  neutrino the parameterization of choice is the same  for
$V_{CKM}$ and $U_{PMNS}$. For $V_{CKM}$ matrix, which is $3\times 3$ matrix in SM,  a
standard parameterization is the following [10]:
\begin{equation}
\left(\begin{array}{ccc}
  c_{12}c_{13} &     s_{12}c_{13} & s_{13}e^{-i\delta}\\
    -s_{12}c_{23}-c_{12}s_{23}s_{13}e^{i\delta}     &
c_{12}c_{23}-s_{12}s_{23}s_{13}e^{i\delta} & s_{23}c_{13}\\                
            s_{12}s_{23}-c_{12}c_{23}s_{13}e^{i\delta}  &
-c_{12}s_{23}-s_{12}c_{23}s_{13}e^{i\delta}  & c_{23}c_{13} 
\end{array}\right) ,
\end{equation}      
where  $c_{ij}  =  cos\theta_{ij},  s_{ij} = sin\theta_{ij}$.  
If  number  of  lepton
generations equal to three and neutrino are Dirac type,  then
one  can  use  the similar parameterization for  $U_{PMNS}$.  When
neutrino  are  Majorana type, the dependence on two  Majorana
phases $\chi_1, \chi_2$  should be taken into account for $U_{PMNS}$. In this
case  an  appropriate form of $U_{PMNS}$  parameterization  is  as
follows:
$$
\left(\begin{array}{ccc}
  c_{12}c_{13} &     s_{12}c_{13} & s_{13}e^{-i\epsilon}\\
    -s_{12}c_{23}-c_{12}s_{23}s_{13}e^{i\epsilon}     &
c_{12}c_{23}-s_{12}s_{23}s_{13}e^{i\epsilon} & s_{23}c_{13}\\                
            s_{12}s_{23}-c_{12}c_{23}s_{13}e^{i\epsilon}  &
-c_{12}s_{23}-s_{12}c_{23}s_{13}e^{i\epsilon}  & c_{23}c_{13} 
\end{array}\right)
\left(\begin{array}{ccc}
  e^{-i\chi_1} &  0 & 0\\
   0 & e^{-i\chi_2} & 0\\                
   0  & 0  & 1 
\end{array}\right) ,
$$
where  $c_{ij}  =  cos\eta_{ij},  s_{ij} = sin\eta_{ij}$. 
Others methods of $V_{CKM}$  and
$U_{PMNS}$ parameterization are possible, among them it should  be
noted  the  Wolfenstein parameterization for $V_{CKM}$ [11],  when
the  experimental  magnitudes of mixing matrix  elements  are
properly accounted for.
\begin{equation}
s_{12} = \lambda,\quad  s_{23} = A\lambda^2, \quad
 s_{13}e^{-i\delta} = A\lambda^3(\rho-i\eta )
\end{equation}
This  parameterization  can  be used  advantageously  in  all
orders  of $\lambda$  even  if effects of new physics beyond  SM  take
place [12, 13].

\smallskip
{\bf  2. Experimental values of quark mixing parameters}

     Let  us consider quark mixing matrix in parameterization
(3), in doing so angles $\theta_{ij}$ specify the mixing between $i-$  and
$j-$generations.  A  qualitative analysis of experimental  data
alone  [1]  point  to  a  number of  interesting  facts.  For
instance,  small  values of angles  $\theta_{13}$ and $\theta_{23}$  tell  us  that
third generation mix weakly with others and at $\theta_{13} =$$\theta_{23}= 0$  it
decouples,  and  only  mixing between first  two  generations
remains  with   $\theta_{12}  =$  $\theta_C$  , where $\theta_C$  is  the  Cabbibo  angle.
Although  this  is  a  rough approximation,  nevertheless  it
frequently used at qualitative description of low energy weak
interaction  processes. It is known from  experimental  data,
that $ cos \theta_{13}$ differs from unity only in the sixth place, thus
the  following  equalities  can be  taken  to  a  quite  good
approximation: $V_{ud} =c_{12}$ , $V_{us} =s_{12}$ , 
$V_{cb} =s_{23}$  ¨ $V_{tb}  =c_{23}$.
Note  that  the  phase $\delta$,  which  associated  directly  with
breaking CP invariance and always appears with the factor $s_{13}$,
  lies  in  the  range $[0, 2\pi)$ close to the  value  $60^{\circ}$  (more
precisely $63^{\circ}$$(+15^{\circ}/-12^{\circ}$) [1]).

     At  present  a  large amount of data is obtained  for  a
determination of matrix element $V_{ud}$ . We have data for  beta-
decays  of a free neutron and a pion on the one hand, on  the
other  data for beta-decays of nuclei. The most precise value
for $V_{ud}$  is found now in super allowed $0^+\to  0^+$ beta-decays  of
light  and  intermediate  nuclei.  In  doing  so  theoretical
results used concerning estimations of matrix elements on the
basis  of hypothesis of the vector current conservation  with
Coulomb  and  nuclear  structure corrections,  together  with
corrections due to exchanges of virtual $\gamma-$quanta and  $Z-$bosons
between  nucleons'  quarks and between nucleons'  quarks  and
final  leptons.  In Ref. [14] the weighted  value  of  twelve
super  allowed $0^+\to  0^+$ beta-decays of nuclei is  given,  which
correlates  well with the weighted value for three processes,
namely,  the super allowed $0^+\to  0^+$ beta-decays of nuclei,  the
beta-decay  of a free neutron and beta-decay of a pion. This
value is equal to
\begin{equation}
                         |V_{ud}|       =      0.9738  \pm     0.0004
\end{equation}
In  the  recent review [1] the most precise value  is  given,
which  obtained only from data for nine super allowed $0^+ \to  0^+$
beta-decays of nuclei:
\begin{equation}
                         |V_{ud}|       =     0.97377 \pm     0.00027
\end{equation}

     In  regard to the values $|V_{ud}|$, extracted from  data  for
beta-decays of a free neutron and a pion, they should be more
free  from uncertainties in comparison with nuclear data,  as
to  they relate to elementary processes. However, things  are
not  so simple. The matter is that even in the principal case
for $V-A-$variant of weak lepton-hadron interaction in order to
extract  unambiguously the data there is a need to carry  out
two  independent experiments as a minimum. For the beta-decay
of neutron as usual one pick out the experiments devoted to a
determination of life time for a neutron and a measurement of
a  spin-electron correlation in decay of a polarized neutron.
In  the  series  of most precise experiments during  90s  the
technique  of ultra cold neutrons (UCN) had been used  (e.g.,
see  [15]). As a result the values for life time of a neutron
$\tau_n$ and  the  A, which is a correlation parameter, have  been
obtained  ($\tau_n$ $ = 885.7\pm 0.8 c$, $A=-0.1161\pm0.0007$). 
These  values yield [14]
\begin{equation}
                         |V_{ud}|  =  0.9741 \pm  0.0020,
\end{equation}
this  is  in  accordance with data for $0^+\to 0^+$  nuclei  beta-
decays.  But  in  the  early   2000s  in  the  UCN  procedure
essentially  new  phenomena have been found, which  connected
with  super  small heating and anomalous interaction  of  UCN
with  container walls. These facts led to a critical analysis
of  previous  results  and a necessity  of  new  methods  for
setting up experiments. Performed experiments for the neutron
life  time  gave  the following value: 
$\tau_n$ $= 878.5\pm0.8  c$  [16],
which differs from the result of Ref. [17], contributing  the
major  portion  in the world averaged value,  more  than  for
magnitude of $5\sigma$. The new value of the neutron life time 
leads to the value:
\begin{equation}
                        |V_{ud}| =   0.9781 \pm  0.0020
\end{equation}
This  value  is  not in a good accordance with the  unitarity
condition for $V_{CKM}$ in view of recent data for the $V_{us}$, it
will  be  discussed below. Thus the question of  the  neutron
life  time is still an open question and further experimental
investigations are needed.

     Some  difficulties  arose  with  a  determination  of  a
magnitude of matrix element $V_{us}$, because of the value 
$|V_{us}|  =
0.2200 \pm 0.0026$ used previously, which has been obtained  from
$K_{e3}$ decays, led to the breaking of the unitarity condition on
the  confidential  level  of $2\sigma$ [14,  15].  This  problem  was
discussed  repeatedly and the assumption that effects  beyond
the  SM  took  place  in the neutron decay  was  one  of  the
possible  solutions.  However, recent  experiments  for  kaon
decays give a new value of $V_{us}$ [14]:
\begin{equation}
                         |V_{us}| =   0.2254 \pm 0.0021,
\end{equation}
which,  as  it  will be seen below, do not violate  the  $V_{CKM}$
unitarity without regard for the result of Ref. [16].

     Really  the  $V_{ub}$  value,  which extract  from  data  for
semileptonic decays $b\to ul\bar\nu_l$ and charge conjugated processes, now
is equal to [1]:
\begin{equation}
                         |V_{ub}| =  (4.31\pm 0.30)\times 10^{-3}
\end{equation}
So  if one take into account the results (6), (9), (10)  the
unitarity condition for $V_{CKM}$ matrix is accurate to $1\sigma$:
\begin{equation}
                 |V_{ud}|^2 + |V_{us}|^2  +  |V_{ub}|^2  =   0.9991(11)
\end{equation}
Using  for $|V_{ud}|$  the value (7) obtained from the neutron  life
time,  one  obtain  an  increase of an  uncertainty  for  the
unitarity condition:
\begin{equation}
                 |V_{ud}|^2+   |V_{us}|^2  + |V_{ub}|^2  =  0.9997(40)
\end{equation}
Whereas using the result of Ref. [16] only we obtain
\begin{equation}
                 |V_{ud}|^2+  |V_{us}|^2  +  |V_{ub}|^2  =  1.0075(40),
\end{equation}
which  lead  to  the  magnitude of right  side  of  Eq.  (14)
exceeding  unity  more than $1\sigma$. It is clear,  that  additional
efforts  are  needed  in  order to resolve  the  experimental
problem  of the neutron life time. At the same time  we  have
the  validity of the unitarity condition (11) for  the  first
row  of $V_{CKM}$ matrix using the world average values, as it was
pointed out above.

     Let  us  go to the next row of the quark mixing  matrix,
which is connected with processes with charm quarks. The  $V_{cd}$
value  is extracted from data for production of charm  quarks
involving neutrino, antineutrino and  $d$ quarks. According  to
the recent data [1] the value is
\begin{equation}
                         |V_{cd}| =      0.230 \pm  0.011
\end{equation}
     Using   along   with  experimental   data   results   of
evaluations  in the frame of the Lattice QCD,  the  following
$|V_{cs}|$  value can be obtained [1]:
\begin{equation}
                       |V_{cs}| =    0.957 \pm 0.017\pm 0.093,
\end{equation}
where the first uncertainty for $|V_{cs}|$  is experimental one, the
second one is the uncertainty of theoretical calculations.

     The  $|V_{cb}|$  value  is  obtained from data  for  exclusive
decays $B\to\bar D^* l^+\nu_l, B\to\bar Dl^+\nu_l$, and for inclusive 
decays  $\bar B\to Xl\bar\nu_l$ with  a
determination  of  a  charge  lepton  characteristics,  which
subsequently are correlated each other. A difference  of  $|V_{cb}|$
values derived from exclusive and inclusive processes lies in
the  range  of  experimental uncertainties.  The $|V_{cb}|$   value
averaged is [1]:
\begin{equation}
                        |V_{cb}| =    (41.6\pm 0.6)\times 10^{-3}  ,
\end{equation}

     Experiments  for $W^{\pm}$ bosons' decays carried out  at  LEP-2
make  possible  to  verify the unitarity condition  for  $V_{CKM}$
matrix for sum $\sum_{u,c,d,s,b} |V_{ij}|^2$ [1]:
\begin{equation}
  |V_{ud}|^2+ |V_{us}|^2 + |V_{ub}|^2+|V_{cd}|^2+ 
|V_{cs}|^2 + |V_{cb}|^2 =  2.002\pm 0.027
\end{equation}
Note  that the unitarity conditions for columns and  rows  of
$V_{CKM}$  matrix  can be represented on a complex  plane  as,  so
called, "unitary triangles".

     Nowadays   there   are  experimental   and   theoretical
difficulties  for determinations of matrix  elements  of  the
third row. For example, usually absolute values for $|V_{td}|$   and
$|V_{ts}|$,  which  is  small enough, ($|V_{td}|$
$ =(7.4\pm 0.8)\times 10^{-3}$,  $|V_{ts}|$
$=(40.6\pm  2.7)\times 10^{-3}$), are determined at the assumption of  unity
absolute  value  for  $V_{tb}$. However,  at  Tevatron  only  the
limitation  $|V_{tb}| > 0.78$ has been obtained with  the  help  of
comparison of data for $t$ quarks decays into $b$ and $s, d$ quarks
[1].  In addition estimations for two-loop contributions  for
decay   width   $\Gamma(Z\to b\bar b)$  give  the  following  value:  
$V_{tb} = 0.77(+0.18/-0.24)$.
     It should be noted that at present main uncertainties of
some  matrix  elements  are due to theoretical  contributions
rather  than  experimental ones. They  are  caused  by  using
various  models  for calculations of matrix elements  on  the
basis  of existing data. With help of models and data,  which
give  the  most precise matrix elements values, the following
ranges for $V_{CKM}$ values have been obtained within $1\sigma$ (they  do
not always symmetric about dominant absolute values) [1-3]:
\begin{equation}
\left(\begin{array}{ccc}
                        0.97360 \div 0.97407 &    0.2262\div 0.2282 &
0.00387 \div 0.00405\\
        0.2261\div0.2281&   0.97272\div0.97320&
0.04141\div0.04231\\
           0.00750\div0.00846& 0.04083\div0.04173& 0.999096\div0.999134
\end{array}\right) ,
\end{equation}      
Then   the  angles $\theta_{ij}$,  entering  in  the  $V_{CKM}$  matrix   in
parameterization (3), are equal to:
\begin{equation}
                        \theta_{12}= 13.14^{\circ}\pm 0.06^{\circ},
     \theta_{23}= 2.43^{\circ}(+0.01^{\circ}/-0.05^{\circ}),             
                        \theta_{13}= 0.23^{\circ}\pm 0.01^{\circ}
\end{equation}

\smallskip
{\bf     3.  Quark  mixing angles $\theta_{ij}$  and masses  of  constituent
quarks}

     It  is known that matrix elements values for $V_{CKM}$   (18)
presented  above  cannot  be  evaluated  from  the  first  SM
principles.    Nevertheless   there   are   a    number    of
phenomenological  approaches,  which  make  it  possible   to
evaluate   these  values.  From  our  standpoint   the   most
interesting  approaches  are those, where  relations  between
mixing  angles  and  quark masses can be found.  One  of  the
approaches of such kind was developed in the Fritzsch's works
[18],  where the connection between mixing angles and current
quark  masses has been established. The authors of Ref.  [19]
modified   this   approach  and  suggest  using   masses   of
constituent quarks instead of current ones. Let us name  this
approach   as  the  Fritzsch-Scadron-Delbourgo-Rupp  approach
(FSDR   approach).  Below  quark  mixing  angles  and   their
uncertainties are evaluated with masses of constituent quarks
obtained in the framework of the relativistic model of  quasi
independent  quarks [20, 21]. Mass values  for  light  quarks
within  uncertainties  of the model of  Refs.  [20,  21]  are
correspond  to  values listed in Refs.  [19,  22],  but  mass
values for heavy quarks turn out somewhat different.

     Relations  obtained in the framework of 
$SU(2)_L\times  SU(2)_R\times
U(1)$  gauge theory form a basis of FSDR approach, the  gauge
theory initially is symmetric with respect to left and  right
currents [18]. The relations between mixing angles and  quark
masses  arise  under assumption that a special  mechanism  of
mass  generation for quarks proceeds like a  cascade.  It  is
suppose  that at first mass values for $c, s, d$ and  $u$  quarks
are  zero,  while  mass values of heavy $b$ and  $t$  quarks  are
initial  parameters of the theory. Then the  mass  generation
for  light quarks proceeds. First masses of $c, s$ quarks  came
via  the  mixing  due  to the weak interaction.  Subsequently
masses  of  $d$  and  $u$  quarks emerged in  an  analogous  way.
Relations between mixing angles and quark masses obtained  by
the consideration of such process [18] later on were modified
and  supplemented in Ref. [19]. In doing this mass values  of
current  quarks,  governing mixing angles in  the  Fritzsch's
model,  were  substituted  with mass  values  of  constituent
quarks.  Moreover the expression for an auxiliary angle $\varphi_{sd}$ was
proposed for a calculation of the Cabbibo angle. The angle $\varphi_{sd}$
can  be expressed in terms of $\pi$ and $K$ decay constants $f_{\pi}$,  $f_K$
and masses of constituent $S$ and $D$ quarks $m_S$, $m_D$:
\begin{equation}                     
 sin2\varphi_{sd} = \frac{2{\sqrt 2}\pi(f_K-f_{\pi})}{\sqrt 3(m_S - m_D)}   
\end{equation}

    Other  two  angles  $\varphi_{cu}$ and $\theta_{23}$  can be calculated
 through formulae, which depend only on quark masses
\begin{equation}                     
 sin2\varphi_{cu} =\sqrt{ \frac{m_S - m_D}{m_C - m_U}}   
\end{equation}
\begin{equation}                     
 sin2\theta_{23} = \sqrt{ \frac{m_C - m_U}{m_T - m_C}}      
\end{equation}                                  
The  angle $\theta_{12}$, which is the Cabbibo angle $\theta_{C}$, 
is equal to the
difference of angles $\varphi_{sd}$ and $\varphi_{cu}$: 
$\theta_{C}  =$ $\varphi_{sd} - \varphi_{cu}$.

     In  Refs.  [19,  22]  the constituent  quark  model  for
hadrons  was employed for a determination of mass  values  of
light  constituent quarks with the help of experimental  data
on  mass  spectra of mesons and baryons, and magnetic moments
of baryons. By this means the following values were obtained:
\begin{equation}                     
     m_U  =  335.5  MeV,\quad  m_D  = 339.5 MeV,\quad   m_S  =  485.7  MeV
\end{equation}
Mass  values  of heavy constituent quarks in Ref.  [19]  were
determined as one-halves mass values for $1^{--}$ quarkonia in their
ground states, hence the following values were used:
\begin{equation}                     
     m_C     =     1550     MeV,  \qquad    m_B    =  4730     MeV
\end{equation}

     This  method of a determination of mass values for heavy
quarks may consider as a quite rough estimation. More precise
mass values for constituent quarks can be found with the help
of  solution  of problem concerning ground states  of  quark-
antiquark  systems  considered in hadrons' potential  models,
for  instance,  in  the relativistic quasi independent  quark
model  with a linear rising confinement potential and a quasi
Coulomb  potential [20, 21]. In the framework of  this  model
the   hypothesis  of  the  universality  of  the  confinement
potential was verified for as heavy as light quarks with  the
coefficient of a slope of a linear rising potential equal  to
$\sigma =  0.20\pm   0.01  GeV^2$. New characteristic  constants  for  a
confinement domain, which have dimensions of mass and length,
can  be  associated  with  the obtained  coefficient $\sigma$ ("the
string tension"): $\mu_C =$$ 0.45\pm0.02 GeV$, $\lambda_C =$
$ 0.44\pm0.02 Fm$. As  this
takes  place,  $\mu_C$  defines typical magnitudes  of  transversal
impulses $ <p_T>$  for quarks-partons inside  hadrons,  while  a
radius of a perturbative domain surrounded a current quark is
equal  to $r_C =$$\lambda_C/2 =$$ 0.22\pm0.01 Fm$. The domain $r > r_C$
 is  most likely  to  be  the domain of a formation for  a  constituent
quark  due  to nonperturbative interactions. Mass values  for
light  constituent  quarks evaluated  in  the  model  are  in
accordance  with  the values (23), however  mass  values  for
heavy quarks differ considerable from the values (24) and are
equal to
\begin{equation}                     
     m_C     =     1610     MeV, \qquad     m_B    =     4950     MeV
\end{equation}
In  the framework of the relativistic quasi independent quark
model  systematic uncertainties can be also estimated for  as
light as heavy quarks.
$$
   m_U = 335\pm2 MeV, \quad m_D = 339\pm2 MeV,\quad m_S = 485\pm8 MeV,
$$
\begin{equation}                     
          m_C    =    1610\pm15    MeV, \qquad  m_B    =    4950\pm20    MeV.
\end{equation}

    Now  we use the results (26), for quark mixing angles  to
be  refined. By using formula (21) one can obtain $sin2\varphi_{cu}  =$
$0.338 \pm  0.009$, so the angle magnitude is $\varphi_{cu} =$
$9.9^{\circ}\pm   0.3^{\circ}$.  We
take  the  value  $20.5\pm  0.2 MeV$ for $f_K - f_{\pi}$,  which  do  not
contradict  to  existing data, then  with  the  help  of  the
relation (20) we obtain $sin 2\varphi_{sd} =$$ 0.720\pm0.041$, that  is  
$\varphi_{sd}$$  = 23.0^{\circ}\pm 1.7^{\circ}$. As the result of these 
calculations we  have  the  Cabbibo angle $\theta_C$:
\begin{equation}                     
 \theta_C   =  \varphi_{sd}  -  \varphi_{cu} =      13.1^{\circ}\pm 1.7^{\circ}
\end{equation}

    With  account  of data [1] the mass of the constituent  $T$
quark  can  be taken as $m_T =$$ 173\pm 3 GeV$, so with the  formula
(22) one can obtain the following value for the $\theta_{23}$:
\begin{equation}                     
    \theta_{23}   =    2.47^{\circ}\pm    0.03^{\circ}
\end{equation}
    As  for  the  $\theta_{13}$  angle  we  use  a  relation,  which  is
immediately apparent from data, namely: $\theta_{13} = \theta_{23}/12$. 
 Thus the $\theta_{13}$ value is equal to
\begin{equation}                     
    \theta_{13}     =     0.206^{\circ}\pm0.003^{\circ}
\end{equation}
    A  comparison  of the quark mixing angles obtained  above
with the experimental values for these angles (see equalities
(19))  shows  that an accordance between these results  takes
place within $3\sigma$ uncertainties.

\smallskip
{\bf     4.  Hypothesis of complementary for quark  and  neutrino
mixing angles and neutrino mixing angles $\eta_{ij}$}

     Let us use quark mixing angles evaluated in the previous
section for checking and refinement of mixing angles  in  the
neutrino  sector  with  the  aid  of  the  hypothesis  of   a
complementary  and equality for mixing angles of  quarks  and
neutrinos [23, 24]. In the framework of the hypothesis it  is
suppose that the angles $\theta_{13}$ are $\eta_{13}$ equal each other, while the
angles $\theta_{12}$ and $\eta_{12}$, $\theta_{23}$ and $\eta_{23}$ complement each other within  the
$\pi/4$ angle, namely:
\begin{equation}                     
                          \theta_{13}  = \eta_{13},\quad
                        \theta_{12} + \eta_{12} = \pi/4,\quad
                                          \theta_{23} + \eta_{23} = \pi/4
\end{equation}

    If  we insert the quark mixing angles $\theta_{ij}$ evaluated in the
previous section, then it is an easy matter to calculate with
formulae  (30)  neutrino mixing angles $\eta_{ij}$,  involved  in  the
$U_{PMNS}$ matrix:
\begin{equation}                     
                         \eta_{12}= 31.9^{\circ}\pm 1.7^{\circ},\quad
   \eta_{23}= 42.53^{\circ}\pm 0.03^{\circ},\quad                        
        \eta_{13}= 0.206^{\circ}\pm 0.003^{\circ}
\end{equation}

It  should be noted that the predicted value of $\eta_{13}$  angle  is
rather  small, so a experimental determination of this  value
is  a  difficult  problem. Of course, the uncertainties shown in Eqs. 
(27), (28), (29) and (31) should be taken as model dependent ones,
so should be increased by unknown systematic (theoretical) uncertainties.
However if  the  hypothesis  of  a
complementary and equality for mixing angles is true, the  $\eta_{13}$
angle is nonzero because of the $\theta_{13}$ is nonzero (see Eq. (29)).
This  fact is principal since only in this case a $CP$ breaking
can   appear  in  processes  involving  neutrinos  as   Dirac
particles.  For  Majorana  neutrinos  $U_{PMNS}$  matrix  contains
additionally  two  phase parameters $\chi_1$ and $\chi_2$,  which  lead  to
observable  effects in some processes, for  instance,  in $0\nu2\beta$
decays of nuclei.

    The  evaluated values for neutrino mixing angles (31) can
be  compared with angles' values found experimentally. It  is
known  that  recent experiments confirm the theoretical  idea
about  oscillations  of electron and  muon  fluxes  in  their
passage of long distance from a source [7-9] on the one hand,
on  the other they provide an explanation for the deficit  of
solar  neutrinos and verify once more the Sun standard  model
[25-27].  Experimental data from KamLAND [28], SNO [29],  K2K
[30], Super-Kamiokande [31] and CHOOZ [32] give the following
restrictions neutrino mixing angles [1]:
\begin{equation}                     
   \theta_{sol} = 34.01^{\circ}(+1.31^{\circ}/-1.56^{\circ}),\quad
       \theta_{atm} >36.78^{\circ},\quad
    \theta_{chz} <  12.92^{\circ}
\end{equation}
In  notations  of our paper $\theta_{sol} = \eta_{12}$, 
$\theta_{atm} =  \eta_{23}$,  $\theta_{chz}  =  \eta_{13}$.
Besides  that an additional restriction for the mixing  angle
$\theta_{chz}$  can  be obtained with the aid of data for neutrino  flux
from  the  star  SN  1987A  [33, 34]  under  assumption  that
neutrino   masses  form  a  normal  hierarchy  [35-37].   The
correlation  of mixing angles (31) and (32) give  us  a  firm
evidence  for  the validity of hypothesis of a  complementary
for mixing angles within existing experimental uncertainties.
As  for  the  equality  of the $\theta_{13}$  and   $\eta_{13}$   angles  then  a
performance  of  experiments are required  for  more  precise
confirmation.   It  is  worth  noting  that  the   calculated
uncertainties  for  $\eta_{12}$, $\eta_{23}$ and $\eta_{13}$ angles  are  considerably
smaller   experimental  ones.  So  the  hypothesis  used   is
consistent  and  can  be applied for independent  estimations
neutrino  mixing angles on the basis of quark  mixing  angles
since the last ones are measured with a high accuracy.

\smallskip
{\bf     5. Conclusions and discussion}

    It  is  generally  recognized that SM is  the  transition
stage  of a more perfect and universal theory notwithstanding
impressive  achievements of the SM. Several  problems  remain
open,  such as experimental verification of the existence  of
Higgs particles, the large number of SM constants, which  are
pure  phenomenological parameters. For  instance,  there  are
experimental  values of quark and lepton  masses,  and  quark
mixing  angles.  The  discovery of neutrino  oscillations  in
fluxes   of   solar,  reactor,  atmospheric  and  accelerator
neutrinos  only  adds  to  the  complexity  of  the   problem
discussed,  because of shows the necessity of increasing  the
number of parameters by reason of neutrino nonzero masses and
mixing  angles. Therefore revealing of relations  between  SM
constants and searching of a more general theory than the SM,
for  instance,  a  GUT  are  urgent problems  of  present-day
investigations.   In  a  GUT  framework   the   gauge   group
$SU(3)_c\times SU(2)_L\times U(1)_Y$ goes over, as a rule, into a gauge  group
with more simpler structure. For one of GUT variants [38, 39]
the group of this type is the $SO(10)$ group with the following
pattern of spontaneous gauge symmetry breaking
\begin{equation}
    SO(10)\to  SU(4)_{ec}\times SU(2)_L\times SU(2)_R
\to  SU(3)_c\times SU(2)_L\times U(1)_Y ,
\end{equation}
where  $SU(4)_{ec}$ is the group of extended color  [40].  In  the
$SO(10)$  model  the  fundamental  spinor  representation  with
dimension  equal  to  $16$  is  filled  by  one  left   fermion
generation consist of quarks, leptons including a $CP$  partner
for  a right neutrino. In the framework of this model nonzero
neutrino  masses  naturally  arise.   This  model  does   not
contradict  the  existing restrictions  associated  with  the
value  of the Weinberg angle, the lifetime of the proton  and
the  neutrino  masses  [1]. Notice that  the  $SU(2)_L\times SU(2)_R$
gauge symmetry is the basis for the FSDR formulae, which  are
applied   in  the present paper. Besides the mass  values  of
constituent quarks and their absolute uncertainties are used,
which have been obtained in the framework of the relativistic
model   for  quasi  independent  quarks  with  the  universal
confinement  potential. As a consequence the accordance  with
the  data for quark mixing angles has been achieved. Moreover
the  neutrino mixing angles have been evaluated with  a  high
accuracy  by  using  the hypothesis of  a  complementary  and
equality for mixing angles of quarks and neutrinos.

     It  should  be  noted, that as distinct from  the  quark
sector  the  independent description  of  a  generation  mass
mechanism  and  mixing do not achieved much success  yet.  At
present it is clear that the neutrino mixing has its peculiar
features  connected with a possibility for  neutrino   be   a
Dirac  or  Majorana particle. In the last case  supplementary
circumstances must be taken into account, when one  transform
flavor  neutrino states to mass neutrino states in  order  to
bring  a  mass  matrix  into  a diagonal  form.  As  analysis
performed  in  the  works [41, 42] in the  framework  of  the
special  two  flavor  Pauli model  shows,  if  neutrinos  are
complex states from Dirac and Majorana components then  their
mixing  can  be determined by a ratio of Dirac  and  Majorana
contributions  in  effective neutrino  masses.  In  its  turn
effective  masses  govern oscillation lengths  of  neutrinos.
From  this point of view, it is possible, that the hypothesis
of  a  complementarity used by us  point to a  similarity  of
mixing   mechanisms  for  quarks  and  neutrinos,  and   this
hypothesis  need  further  consideration   with  account   of
peculiarities for quark and neutrino sectors. So tackling the
question  about a Dirac or Majorana nature of  neutrinos  has
the  principal  significance.  It  is  known,  that  the  key
experiment is a discovery of the neutrinoless mode for double
beta nuclear decays. An observation of this mode signals that
a  neutrino  has  Majorana  properties,  while  an  effective
coherent mass
\begin{equation}
 m_{\beta\beta}  =  U^2_{e1}m_1  +  U^2_{e2}m_2  +  U^2_{e3}m_3  ,
\end{equation}
measured  in  experiments of such kind, depends  on  Majorana
phases  entering into $U^2_{ei}$, $i=1, 2, 3$. The current data  give
for the $m_{\beta\beta}$   value the following limitation [43-46]:
\begin{equation}
                |m_{\beta\beta}|  <   0.55  eV
\end{equation}
Moreover,  direct measurements of neutrino mass in  the  beta
decay of tritium [47, 48] lead to
\begin{equation}
                         m_{\nu}  <     2.2    eV
\end{equation}
At  the same time oscillation data show the occurrence of the
lower limitation for one of neutrino masses [49]:
\begin{equation}
                       m_{\nu}  >8.5 \cdot  10^{ -3} eV
\end{equation}
Hence  the  question about the character  of  mixing  in  the
lepton sector is still open.

     In  conclusion  the  results of the work  indicate  that
mixing  angles for quarks and leptons are not independent  SM
parameters,  supposedly  they  relate  to  masses  of   these
particles.  Moreover  mechanisms of mass  generation  in  the
quark and lepton sectors ought to be dependent each other  as
indicated  by correlations between quark and neutrino  mixing
angles. Undeniably the major importance for the resolution of
the  mixing problem, primarily in the lepton sector  has  got
and will be got results of experiments on neutrinoless double
beta nuclear decays.

     The work is partially supported by the grant \# 27 of the
RRC  "Kurchatov Institute" on fundamental researches in  2006
year.

\end{document}